\documentclass[sigconf]{acmart}
\usepackage{amsmath}
\usepackage[acronym]{glossaries}
\usepackage{subcaption}
\usepackage{enumitem}

\newcommand{\circnum}[1]{\raisebox{.5pt}{\textcircled{\raisebox{-.9pt}{#1}}}}

\newacronym{AI}{AI}{Artificial Intelligence}
\newacronym{GIEOAS}{GIEOAS}{Green In\-fras\-truc\-ture Eval\-u\-a\-tion, Op\-ti\-miza\-tion, and Align\-ment System}
\newacronym{IaC}{IaC}{Infrastructure-as-Code}
\newacronym{WHO}{WHO}{World Health Organization}
\newacronym{SOTA}{SOTA}{State-of-the-Art}
\newacronym{VM}{VM}{Vir\-tual Ma\-chines}
\newacronym{ETG}{ETG}{Enterprise Topology Graph}
\newacronym{EAM}{EAM}{Enterprise Architecture Management}
\newacronym{EA}{EA}{Enterprise Architecture}
\newacronym{GHG}{GHG}{Greenhouse Gas}
\newacronym{ICT}{ICT}{Information and Communication Technology}
\newacronym{DSL}{DSL}{Domain-Specific Language}
\newacronym{SLA}{SLA}{Service Level Agreement}
\newacronym{MOF}{MOF}{Meta-Object Facility}
\newacronym{EMOF}{EMOF}{Essential Meta-Object Facility}
\newacronym{EMF}{EMF}{Eclipse Modeling Framework}
\newacronym{DOML}{DOML}{DevOps Modeling Language}
\newacronym{PG}{PG}{Property Graph}
\newacronym{IPG}{IPG}{Infrastructure Property Graph}
\newacronym{aIPG}{aIPG}{Abstract Infrastructure Property Graph}
\newacronym{LCG}{LCG}{Least Common Generalization}
\newacronym{DRL}{DRL}{Deep Reinforcement Learning}
\newacronym{RL}{RL}{Reinforcement Learning}
\newacronym{IOPS}{IOPS}{Input/Output Operations Per Second}
\newacronym{GPI}{GPI}{Green Performance Indicator}
\newacronym{OS}{OS}{Operating System}
\newacronym{EDMM}{EDMM}{Essential Deployment Metamodel}
\newacronym{ICG}{ICG}{Infrastructure Code Generator}
\newacronym{GSF}{GSF}{Green Software Foundation}
\newacronym{GEL}{GEL}{Green Evaluation Language}

\acmYear{2026}
\acmDOI{XXXXXXX.XXXXXXX}
\acmConference[Greens'26]{10th International Workshop
on Green and Sustainable Software}{Tuesday, April 14 2026}{Rio de Janeiro, Brazil}
\acmISBN{978-1-4503-XXXX-X/2026/04}

\begin{document}

\title{METRION: A Framework for Accurate Software Energy Measurement}

\author{Benjamin Weigell}
\email{benjamin.weigell@uni-a.de}
\orcid{1234-5678-9012}
\affiliation{%
  \institution{University of Augsburg}
  \city{Augsburg}
  \country{Germany}
}

\author{Simon Hornung}
\email{simon.hornung@uni-a.de}
\affiliation{%
  \institution{University of Augsburg}
    \city{Augsburg}
  \country{Germany}}

\author{Bernhard Bauer}
\email{bernhard.bauer@uni-a.de}
\affiliation{%
  \institution{University of Augsburg}
    \city{Augsburg}
  \country{Germany}}

\renewcommand{\shortauthors}{B. Weigell et al.}

\begin{abstract}
The Information and Communication Technology sector accounted for approximately 1.4\% of global greenhouse gas emissions and 4\% of the world’s electricity consumption in 2020, with both expected to rise. To reduce this environmental impact, optimization strategies are employed to reduce energy consumption at the IT infrastructure and application levels. However, effective optimization requires, firstly, the identification of major energy consumers and, secondly, the ability to quantify whether an optimization has achieved the intended energy savings. Accurate determination of application-level energy consumption is thus essential.
Therefore, we introduce an energy attribution model that quantifies the energy consumption of applications on CPU and DRAM at the thread level, considering the influence of Simultaneous Multithreading, frequency scaling, multi-socket architectures, and Non-Uniform Memory Access. To ensure cross-platform applicability, we integrate the proposed model into an extensible framework, METRION, including a platform-independent data model and an initial implementation for Linux systems using Intel CPUs. We evaluate METRION across three different workloads and demonstrate that the energy attribution model can accurately capture the CPU energy consumption of applications targeting solely the CPU with a Mean Absolute Percentage Error of 4.2\%, and the DRAM energy consumption of applications targeting DRAM with an 16.1\% error.
\end{abstract}

%



\maketitle

\section{Introduction}
Increasing digitalization boosts the importance of the \acrfull{ICT} sector, which supports areas from critical infrastructure to personal entertainment. However, this increase comes at a significant environmental cost. In 2020, the \acrfull{GHG} emissions of the \acrshort{ICT} sector accounted for approximately 1.4\% of global emissions and consumed around 4\% of the world’s electricity \cite{Malmodin.2024}. Moreover, it is predicted that by 2030, the ICT sector could consume up to 21\% of global energy, with data centers contributing nearly one-third \cite{Andrae.2015} . These estimates likely understate future consumption, as they do not account for the rising energy demands of AI and blockchain. While data centers are the visible energy consumers, the underlying cause lies in the software applications executed on these infrastructures. 

Therefore, improving energy efficiency requires accurately determining the energy consumption of a software application. Such visibility enables the identification and optimization of energy hot\-spots at the macro- and micro-architectural levels. At the macro-architectural level, energy visibility enables the identification of energy-intensive applications and the development of strategies for reducing overall energy consumption, for example, by shutting down unused applications. At the micro-architectural level, detailed and accurate energy consumption allows fine-grained comparisons between different implementations. For instance, developers can evaluate whether reimplementing a component in a different programming language or applying specific code optimizations lead to measurable reductions in energy consumption \cite{Georgiou.2017}.

To determine the energy consumption, recent hardware offers on-device measurement sensors that measure the energy consumption of the CPU and memory \cite{Khan.2018}. Moreover, RAPL measurements have been shown to correlate strongly with total system power consumption \cite{Khan.2018}. However, these measurements capture energy consumption at the component level and do not directly attribute it to a software application. To overcome this limitation, several attribution approaches combine runtime information with on-device energy measurements to infer application-level energy use.  

Existing methods fall roughly into two categories. Utilization-based models \cite{Petit.2023, CodeCarbon, He.2024}, quantify energy consumption from coarse-grained resource utilization metrics, e.g., CPU time. However, they typically neglect hardware-level effects such as frequen\-cy scaling and Simultaneous Multithreading (SMT), which substantially influence energy behavior \cite{Brondolin.2018, Zhai.2014}. Event-based models \cite{Zhai.2014, Brondolin.2018, Fieni.2020}, in contrast, use Hardware Performance Counters (HWPCs) that correlate closely with CPU power consumption \cite{Bellosa.2000, Bircher.2012}. While these approaches capture fine-grained thread-level detail, they often compromise transparency or completeness. Some approaches model CPU energy consumption and account for SMT effects, but neglect frequency variation and DRAM energy \cite{Zhai.2014, Brondolin.2018}. Other dynamically select HWPC events and builds frequency-specific regression models, but neglecting SMT and Non-Uniform Memory Access (NUMA) effects \cite{Fieni.2020}. Overall, existing models remain limited in scope and do not account for the combined effects of frequency scaling, SMT, and NUMA in multi-socket systems, thereby limiting their accuracy.

Accordingly, we want to answer this research question (RQ): \textit{How can the energy consumption of software applications be accurately and transparently attributed to underlying hardware components, such as the CPU and memory, in complex multi-socket and SMT environments, using only the measurements available from on-device sensors?}
To answer this research question, we present the following contributions:
\begin{enumerate}
\item An interpretable thread-level energy attribution model that determines CPU and DRAM energy using HWPCs, that explicitly accounts for multi-socket topology, NUMA locality, frequency scaling, and SMT, with final aggregation at the application level.
\item METRION, a modular, extensible, and platform-agnostic \linebreak framework for realizing the energy attribution model.
\item A platform-agnostic and extensible data model for platform-independent application of the energy attribution model.
\end{enumerate}

\section{Related Work}
On the one hand, one of the first approaches to determine the energy consumption of software applications is introduced by Flinn et al. \cite{Flinn.1999} with their tool PowerScope. PowerScope correlates external power measurements with kernel events to map energy consumption to software components. 
On the other hand, one of the first HWPC-based approaches is JouleWatcher by Bellosa \cite{Bellosa.2000}, \linebreak which determines the energy consumption of a thread based on HWPC events. The approach requires an initial calibration step in which the energy per event is determined. Bellosa \cite{Bellosa.2000} demonstrates that events such as retired micro-operations per second, floating-point operations per second, second-level cache address strobes, and memory bus transactions correlate with system energy consumption. Do et al. \cite{Do.2009} introduce pTop, one of the early solely software-based tools for determining the energy consumption of applications. pTop determines the energy consumption of an  application by correlating predefined power values of the CPU, hard disk and network interface card (NIC) with the measured resource utilization of these components by the application. In this utilization-oriented line of work, Noureddine et al. \cite{Noureddine.2012} introduce PowerAPI, a purely software-based framework that calculates the power consumption of applications based on their CPU and NIC usage through different power models. Around the same time, Bircher and John \cite{Bircher.2012} propose a software-based approach to determine the power consumption of a complete system using HWPCs. They de\-monstrate that events such as Cycles, Fetched µops, and L3 cache load misses, can be used to build linear regression models that calculate the power consumption of the CPU and memory, respectively. 

Intel introduced, around 2012 with the Sandy Bridge architecture, the Running Average Power Limit (RAPL) interface \cite{Weaver.2012}. Depending on the chip’s architecture, RAPL exposes energy consumption data for different power domains of the processor. For example, the Package domain includes the core and uncore components of the chip \cite{Khan.2018}, while the DRAM domain reports the energy consumption of the attached integrated memory controller. With the introduction of RAPL two kinds of energy granularity attribution tools have emerged \cite{Jay.2023}, software that determines the energy consumption of the whole system, such as Carbon Tracker \cite{Anthony.2020}, CodeCarbon \cite{CodeCarbon}, and Experiment Impact Tracker \cite{ImpactTracker}. And energy measurement software that determines the energy consumption of a single application, such as HaPPy \cite{Zhai.2014}, DEEP-mon \cite{Brondolin.2018}, a new version of PowerAPI \cite{Fieni.2020}, Scaphandre \cite{Petit.2023} and EnergAt \cite{He.2024}, HaPPy \cite{Zhai.2014} uses RAPL energy readings as input for a hyperthread-aware Energy Attribution Model (EAM) to determine the energy consumption of an application. This model leverages non-halted CPU cycles together with hyperthreading information to determine the share of energy consumption for each application. DEEP-mon \cite{Brondolin.2018} refines the approach of HaPPy by making the hyperthread-aware energy attribution resilient to time-shared CPUs and extending its coverage to application containers. Furthermore, Fieni et al. \cite{Fieni.2020} replace the EAM of PowerAPI with an regression-based one that uses RAPL counters and dynamically selects HWPCs to build a frequency-aware EAM. This model determines the energy consumption of an application on both the CPU and DRAM, however primarily targets containers. Scaphandre \cite{Petit.2023} determines the energy consumption of an application by attributing the CPU energy consumption reported by RAPL according to its time share on the CPU. With EnergAt Hè et al. \cite{He.2024} propose an EAM that determines the energy consumption at a thread level while considering influencing factors such as multi-socket NUMA architectures, the noisy-neighbour effect and and multi-tenancy. In summary, existing approaches remain limited in their scope because they do not take into account the combined effects of frequency scaling, SMT, and NUMA in multi-socket systems, thereby limiting their accuracy.

\section{Energy Attribution Model}

Given the increasing availability of on-device sensor hardware that reports the current energy consumption of CPU and memory components, we present an Energy Attribution Model (EAM) that redistributes these component-level energy measurements to threads based on HWPC events and thereby determines the energy consumption of a software application at the thread level.

\subsection{Design Considerations}
The design of the EAM is influenced by hardware and execution characteristics that affect the energy consumption of the application and are discussed in the following as key design considerations.

\subsubsection{Idle and Active Power Consumption} 
The power consumption of a hardware component consists of idle and active parts. Idle power represents the baseline consumption independent of resource usage, while active power depends on actual resource usage. An EAM should attribute active energy to applications based on their resource usage and distribute this idle baseline energy fairly among all running applications.

For example, a CPU with a total power consumption of $50 W$ under load, of which $10 W$ is idle power, running two processes A and B using $10\%$ and $20\%$ of the CPU resources, respectively. If idle power is ignored, the total power ($50 W$) may be distributed proportionally to utilization, resulting in $16.7 W$ for process A and $33.3 W$ for process B. In contrast, when idle power is considered, the active portion ($50 W – 10 W = 40 W$) is first distributed according to utilization: process A consumes 13.3 W and process B $26.7 W$. The idle power is then divided equally between the two ($5 W$ each), leading to total allocations of $18.3 W$ and $31.7 W$, respectively an $8.7\%$ and $5.0\%$ difference. Therefore, EAMs should first isolate active power and then distribute idle power fairly among consumers.

\subsubsection{Multiple-Socket Architectures}
Multiple-socket architectures, in which each socket comprises a CPU package and an associated memory domain, are commonly employed in modern server hardware \cite{He.2024}. However, most current power models do not consider the resulting NUMA effect and therefore incorrectly distribute the energy consumed by applications \cite{He.2024}. This aspect becomes increasingly important as current on-device sensors, such as Intel RAPL, provide energy readings for each CPU package and its attached memory independently. Therefore, an EAM must consider where the applications consume energy. 

\subsubsection{Thread-Level Execution and SMT}
Modern applications and operating systems employ both multi-processing and \linebreak multi-threading. An application can consist of multiple processes, each containing one or more threads. In modern operating systems, the thread is the basic unit of execution and can be viewed as a lightweight process \cite{tanenbaum2006}. The kernel schedules these individual threads on a CPU core. Additionally, SMT enables two or more threads, depending on the processor vendor and model, to execute concurrently on the same physical core. In the case of SMT with factor two, two threads can execute concurrently on the logical cores of the same physical core, as illustrated in Figure \ref{fig:smt_concurrent}. However, this concurrency affects energy consumption, as concurrent execution has been shown to increase total energy usage by approximately 15 \% when using SMT with a factor of two, compared to single-threaded execution \cite{Brondolin.2018}.

\subsubsection{Frequency Effects} 
The power consumption of the CPU is approximated by the CMOS circuit power model, whereby it consists of $A$ the switching activity, $C$ the physical capacitance, $V$ the supply voltage, and $f$ for clock frequency \cite{Dayarathna.2015} and is given by: $P = A \cdot C \cdot V^2 \cdot f$. Furthermore, the supply voltage scales approximately linearly with frequency, so that $P \approx A \cdot C \cdot f^2$ applies \cite{Murali.2006}. Hence, an EAM attributing CPU power consumption to an application should consider the frequency. 

\subsection{Model Foundations}
Before presenting the model itself, we first formalize its foundations by defining how components and their active and idle energy are represented, as well as how an application is modeled.

\subsubsection{Components and their Active and Idle Energy}
For a given device, we consider each hardware component that provides independent energy measurements via on-device sensors, such as Intel RAPL. The current model focuses on the set of CPU packages $P$, where each package consists of a set of logical cores, and DRAM components $D$, which together form the set of considered components $C$:

\begin{equation}
\label{eq:components}
\begin{split}
C &= P \cup D \\
  &= \{ CPU_1, \dots, CPU_n,\} \cup \{ DRAM_1, \dots, DRAM_m \} \quad n, m \in \mathbb{N}
\end{split}
\end{equation}

For each component $c \in C$, the total, active, and idle energy measured over the fixed time interval $\Delta t$ are denoted by $E^{c}_{total}$, $E^{c}_{active}$, and $E^{c}_{idle}$, respectively. Subtracting the idle energy from the total energy, results in the active energy, as shown in Equation (\ref{eq:c_active}). The on-device sensors provide $E^{c}_{total}$ for each component $c$ during the interval $\Delta t$, as well as $E^{c}_{idle}$ under an appropriate configuration.

\begin{equation}
	\label{eq:c_active}
	E^{c}_{active} = E^{c}_{total} - E^{c}_{idle}
\end{equation}

\subsubsection{Dividing Applications into Threads}
To determine the energy consumption of an application $a_i$, the energy attribution model considers each application as a set of threads $t^{a_i}_j$ that belong to the application and are executed during the time interval $\Delta t$:

\begin{equation}
	\label{eq:applications}
	a_i =\{t^{a_i}_1, t^{a_i}_2, \dots t^{a_i}_l\} \quad l \in \mathbb{N}
\end{equation}

Furthermore, a thread is scheduled multiple times on a core for execution, and each scheduling-in and -out of a thread on a core is captured by an execution interval $e^{t^{a_i}_j}_k$. Each execution interval forms a triple containing the core, the scheduling-in timestamp, and the scheduling-out timestamp. For a thread $t^{a_i}_j$, the set $e^{t^{a_i}_j}$ contains all of the thread’s execution intervals $e^{t^{a_i}_j}_k$:

\begin{equation}
	e^{t^{a_i}_j} = \{e^{t^{a_i}_j}_1, e^{t^{a_i}_j}_2, \dots e^{t^{a_i}_j}_p\} \quad p \in \mathbb{N}
\end{equation}

The duration of an execution interval \( e^{t^{a_i}_j}_k \) is given by the duration function \( \delta(e^{t^{a_i}_j}_k) \), and the CPU package to which the core belongs is determined by the location function \( L(e^{t^{a_i}_j}_k, c) \). The location function evaluates to 1 if the core on which the thread is executed during the interval belongs to CPU package \( c \), and 0 otherwise.

Furthermore, to account for SMT, the execution interval $e^{t^{a_i}_j}_k$ of a thread $t^{a_i}_j$ is split into sub-intervals. If the thread $t^{a_i}_j$ is executed concurrently with other threads on the same physical core, the execution interval is divided into consecutive sub-intervals during which it runs either exclusively or concurrently with other simultaneously running threads. Otherwise, if the thread is executed exclusively on a single physical core, the execution interval $e^{t^{a_i}_j}_k$ remains unsplit. The final set of execution intervals for a thread $t^{a_i}_j$ is defined as: 

\begin{equation}
	e^{t^{a_i}_j}_k = \{e^{t^{a_i}_j}_{k,1}, e^{t^{a_i}_j}_{k,2}, \dots, e^{t^{a_i}_j}_{k,s}\} \quad s \in \mathbb{N} 
\end{equation}

\begin{figure}[t]
    \centering
    \begin{subfigure}[t]{0.47\linewidth}
        \centering
        \includegraphics[width=\linewidth]{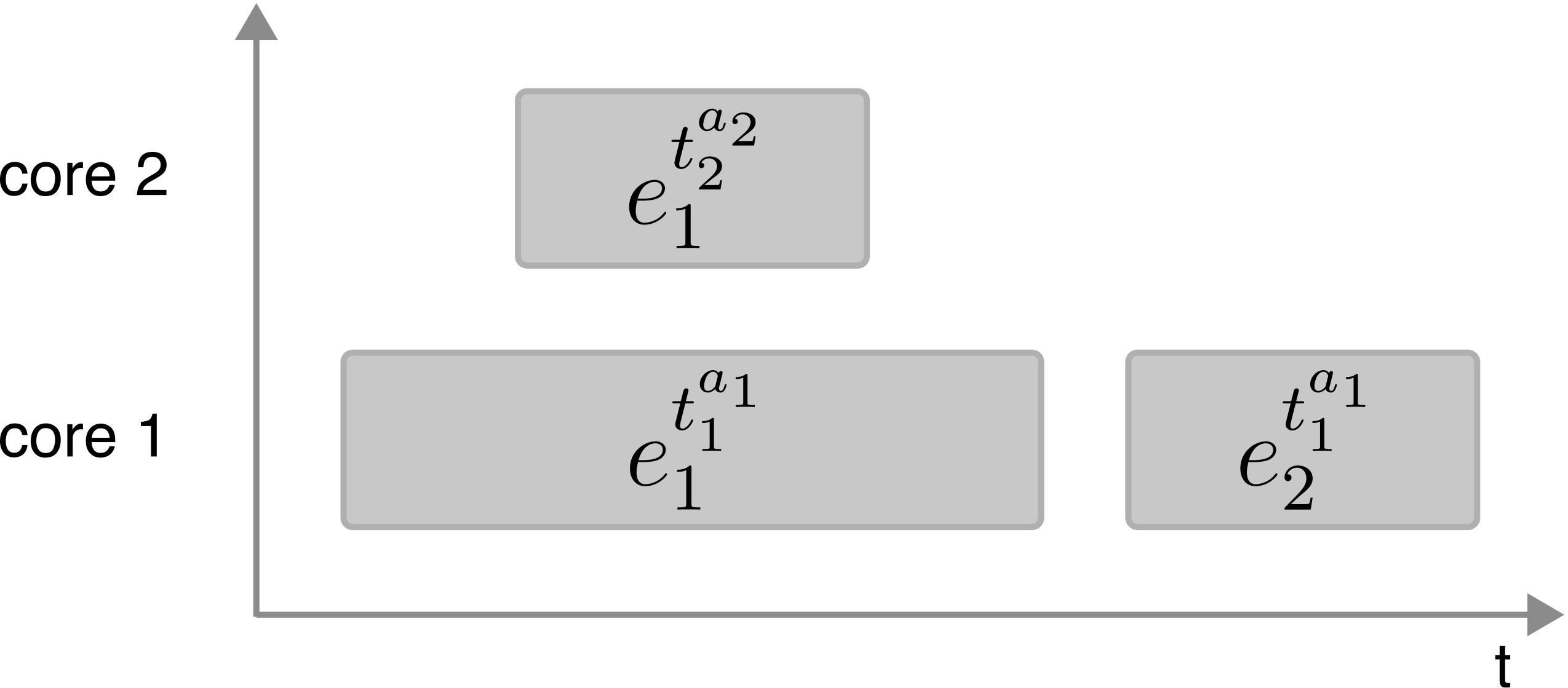}
        \caption{Parallel thread execution.}
        \label{fig:smt_concurrent}
    \end{subfigure}
    \hfill
    \begin{subfigure}[t]{0.47\linewidth}
        \centering
        \includegraphics[width=\linewidth]{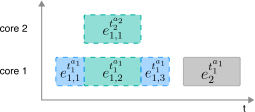}
        \caption{Execution interval splitting.}
        \label{fig:smt_split}
    \end{subfigure}
    \caption{Concurrent thread execution and execution interval splitting of two threads}
    \label{fig:smt_intervals}
\end{figure}

An example of this splitting is illustrated for two threads $t^{a_1}_1$ and $t^{a_2}_2$, with three execution intervals $e^{t^{a_1}_1}_1, e^{t^{a_1}_1}_2$ and $e^{t^{a_2}_2}_1$. The intervals $e^{t^{a_1}_1}_1$ and $e^{t^{a_2}_2}_1$ are executed concurrently, resulting in the sub-intervals $e^{t^{a_1}_1}_{1,1}, e^{t^{a_1}_1}_{1,2}, e^{t^{a_1}_1}_{1,3}$, as illustrated in Figure \ref{fig:smt_split}.

\subsection{The Energy Attribution Model}
The energy consumption of an application $a_i$ is the sum of its active and idle energy shares for each component, as shown in Equation (\ref{eq:application_energy}), thus addressing the first two considerations. The active and idle energy shares of an application are determined from the energy consumption of its threads, which in turn are computed over their respective execution intervals, as described in Equation (\ref{eq:application_active_decomp}).

\begin{equation}
\label{eq:application_energy}
E_{total}(a_i)
  = \sum_{c \in C} \big( E^{c}_{active}(a_i) + E^{c}_{idle}(a_i) \big)
\end{equation}

\begin{equation}
\label{eq:application_active_decomp}
\begin{aligned}
E^{c}_{active}(a_i)
  &= \sum_{j,k,l} E^{c}_{active}\big(e^{t^{a_i}_j}_{k,l}\big),
  \qquad
E^{c}_{idle}(a_i)
  = \sum_{j,k,l} E^{c}_{idle}\big(e^{t^{a_i}_j}_{k,l}\big)
\end{aligned}
\end{equation}

\subsubsection{Active Energy Attribution}
The active energy consumption of a thread during an execution interval, $E^{c}_{active}(e^{t^{a_i}_j}_{k,l})$, is computed independently for each component $c$, and the total active energy of the thread is obtained by summing these per-component contributions. In both cases, the energy consumption depends on the amount of work the thread performs during this interval on a component $c$, relative to the total work performed on that component in the same interval. To capture this, we define component-specific work functions $w_{c}(\cdot)$ that quantify the activity of a thread on component $c$, e.g., $w_{c} = w_{CPU}$ for computational activity, when $c \in P$, and $w_{c} = w_{DRAM}$ for memory activity, when $c \in D$. Based on these work functions, the active energy of component $c$ attributed to an execution interval is given by Equation (\ref{eq:active_energy_general}):

\begin{equation}
\label{eq:active_energy_general}
E^{c}_{active}(e^{t^{a_i}_j}_{k,l}) =
\dfrac{w_{c}(e^{t^{a_i}_j}_{k,l}, c)}%
{\sum_{i',j',k',l'} w_{c}(e^{t^{a_{i'}}_{j'}}_{{k'},{l'}}, c)} 
\cdot E^{c}_{active}
\end{equation}

\paragraph{CPU Work Function} 
The CPU work function quantifies the \linebreak amount of CPU work performed during an execution interval. \linebreak Thereby, the work function respects the design considerations of multiple-socket architectures, thread-level execution and SMT, and frequency effects. 

Each thread executes on a specific logical core within a CPU package $c$. To capture the general work a thread performs during an execution interval, the HWPC Unhalted Core Cycles per Thread ($UCC$) is used. The HWPC $UCC$ captures the number of core cycles executed by a logical core up to a given timestamp, and the difference, $\Delta UCC$, between two timestamps strongly correlates with the CPU's energy consumption during that interval \cite{Bellosa.2000, Bircher.2012, Fieni.2020, Zhai.2014, Brondolin.2018}. However, to account for frequency scaling effects, $\Delta UCC$ is scaled with the core's average frequency ratio during this time interval. To calculate the average frequency ratio, the HWPCs $APERF$ and $MPERF$ are used.  $APERF$ and $MPERF$ measure actual and maximum performance clock counts per logical core, respectively. The HWPC $APERF$ changes with the current operating frequency. In contrast, $MPERF$ increments at the logical core’s reference frequency. The frequency ratio $\frac{\Delta APERF}{\Delta MPERF}$ is greater than $1$ if the logical core operates above the base frequency, such as with turbo boost. It falls below $1$ when the core operates below the base frequency, for example, under energy-saving conditions. Scaling the work by this ratio adjusts it to match the effective operating frequency. 

To consider SMT effects, $\Delta UCC$  is scaled with an SMT factor that is determined by the $\sigma$ function. The $\sigma$ function equals $1$ if the thread executes exclusively on the physical core, and $1.15$ when SMT is active during the interval \cite{Brondolin.2018}. Returning to the previous example, the execution intervals $e^{t^{a_1}_1}_{1,2}$ and $e^{t^{a_2}_2}_{1,1}$ are SMT-active phases and are scaled with a factor of $1.15$, while the other intervals represent exclusive execution and are scaled with $1$, as illustrated in Figure \ref{fig:smt_split}, where the green intervals indicate SMT-active phases.

Overall, the work performed by a thread during the execution time interval on the given CPU package $c$ is specified by Equation (\ref{eq:work_cpu}), whereby the previously defined location function $L(\cdot)$ ensures that only the work performed on the corresponding CPU package contributes to the result:

\begin{equation}
\label{eq:work_cpu}
	w_{CPU}(e^{t^{a_i}_j}_k, c) = \Delta UCC \cdot \frac{\Delta APERF}{\Delta MPERF} \cdot \sigma(e^{t^{a_i}_j}_k) \cdot L(e^{t^{a_i}_j}_k, c)
\end{equation}

\paragraph{DRAM Work Function}
The DRAM work function quantifies a thread’s memory activity, which correlates with energy consumption, during an execution interval, while accounting for the design considerations of multi-socket architectures and thread-level execution. The energy-relevant memory activity of a thread during an execution interval is quantified by the number of off-chip memory transactions it causes that are served by a DRAM component. Off-chip transactions include read and write operations, but only read transactions are modelled, as write operations are first written to the cache and later transferred to the DRAM during cache line eviction, which makes them difficult to attribute to a specific execution interval. The read transactions are recorded by a HWPC that counts DRAM-serviced LLC-miss loads per DRAM component. The change in this counter for DRAM component $d$ during an interval is given by $\Delta O_d(e)$. Further, to account for higher energy consumption due to remote memory access, remotes are scaled with the weighting factor $\gamma$. For local DRAM accesses, $\gamma = 1$, and for remote accesses, $\gamma = 9.67$, as DRAM energy consumption increases by this factor for memory-intensive applications due to remote access \cite{Yu.2017}. Accordingly, the work performed by a thread on a DRAM component during an execution interval is given by Equation (\ref{eq:work_dram}).

\begin{equation}
\label{eq:work_dram}
w_{DRAM}(e^{t^{a_i}_j}_k, d) = \Delta O_d(e^{t^{a_i}_j}_k) \cdot \gamma
\end{equation}

\subsubsection{Idle Energy Attribution} 
The idle energy attributed to an execution interval of a thread is determined by proportionally assigning the idle energy of the component $c$ measured during the monitoring interval $\Delta t$ to that interval. For the CPU, it is attributed according to the thread’s time share on the component $c$ during $\Delta t$, and for the DRAM, it is distributed in proportion to the number of concurrently active threads on the component $c$ during the same monitoring interval, as shown in Equation (\ref{eq:idle_energy_general}).

\begin{equation}
\label{eq:idle_energy_general}
E^{c}_{idle}(e^{t^{a_i}_j}_{k,l}) =
\begin{cases}
\dfrac{
\delta(e^{t^{a_i}_j}_{k,l}) \cdot L(e^{t^{a_i}_j}_{k,l}, c)
}{
\sum_{i',j',k',l'} \delta(e^{t^{a_{i'}}_{j'}}_{{k'},{l'}}) \cdot L(e^{t^{a_{i'}}_{j'}}_{{k'},{l'}}, c)
} \cdot E^{c}_{idle}, & \text{if } c \in P, \\[2.6em]
\dfrac{
1
}{
n(\Delta t, c)
} \cdot E^{c}_{idle}, & \text{if } c \in D.
\end{cases}
\end{equation}

With $n(\Delta t, c)$ representing the number of active threads observed on component $c$ during the monitoring interval $\Delta t$.

\section{METRION}
The proposed energy attribution model is implemented within the METRION framework. The main objective of the framework is to determine the energy consumption of applications on different platforms using the proposed energy attribution model. However, \linebreak achieving platform independence is challenging due to the diversity and continuous evolution of operating systems and hardware. The following section describes how this challenge is addressed within the architecture of the framework and by an extensible, platform-independent data model. In addition, an initial implementation is provided for Linux-based systems with Intel processors.

\subsection{Platform Independent Data Model}
\begin{figure}[h]
\centering
\includegraphics[width=\linewidth]{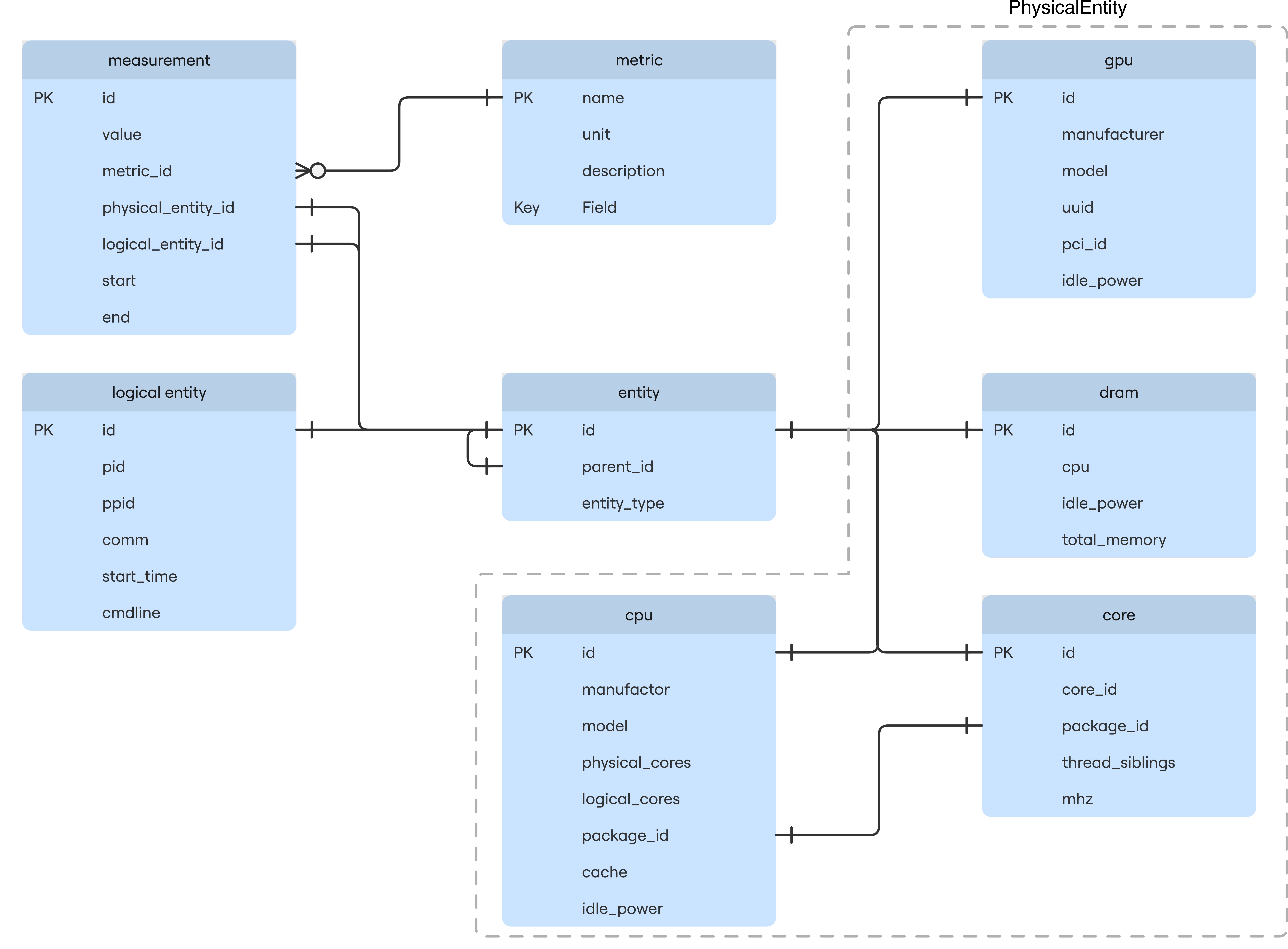}
\caption{The Platform Independent Data Model}
\label{fig:DataModel}
\end{figure}

To ensure platform independence, all components of the METRION framework exchange data through the platform-independent data model (PIDM). The PIDM provides a standardized representation of the required data and abstracts hardware- and operating-system-specific details and thereby enables the framework to operate consistently across platforms. The model is visualised in Figure \ref{fig:DataModel}

The PIDM is structured into the four main parts: Metric, LogicalEntity, PhysicalEntity and Measurement. The PhysicalEntity set represents the component set $C$ of the energy attribution model and includes the physical components such as CPUs, Cores, and DRAM banks along with their minimal platform-agnostic metadata information; also, the model is prepared to support GPUs, though not yet integrated. The LogicalEntity models  the application set $A$ and their corresponding threads. Next, the metric defines what is captured, for example $APERF$ or $E^{c}_{IDLE}$. Finally, the Measurement expresses where, what, when, and how much is measured by linking these elements together: it references the PhysicalEntity on which the event occurred, the Metric describing what was measured, the execution interval via start and end timestamps, and, if applicable, the LogicalEntity for the involved thread. For example, to express $\Delta UCC$ for an execution interval, the following measurement would be taken. The measurement references $t_{j}^{a_i}$ through the logical\_entity\_id, identifies the corresponding core via the physical\_entity\_id, defines the execution interval using the start and stop timestamps, and stores the measured $\Delta UCC$ in the value attribute while referencing the Metric UCC to indicate that the value represents unhalted core cycles per thread.

\subsection{Architecture and Dataflow}
\begin{figure}[h]
\centering
\includegraphics[width=\linewidth]{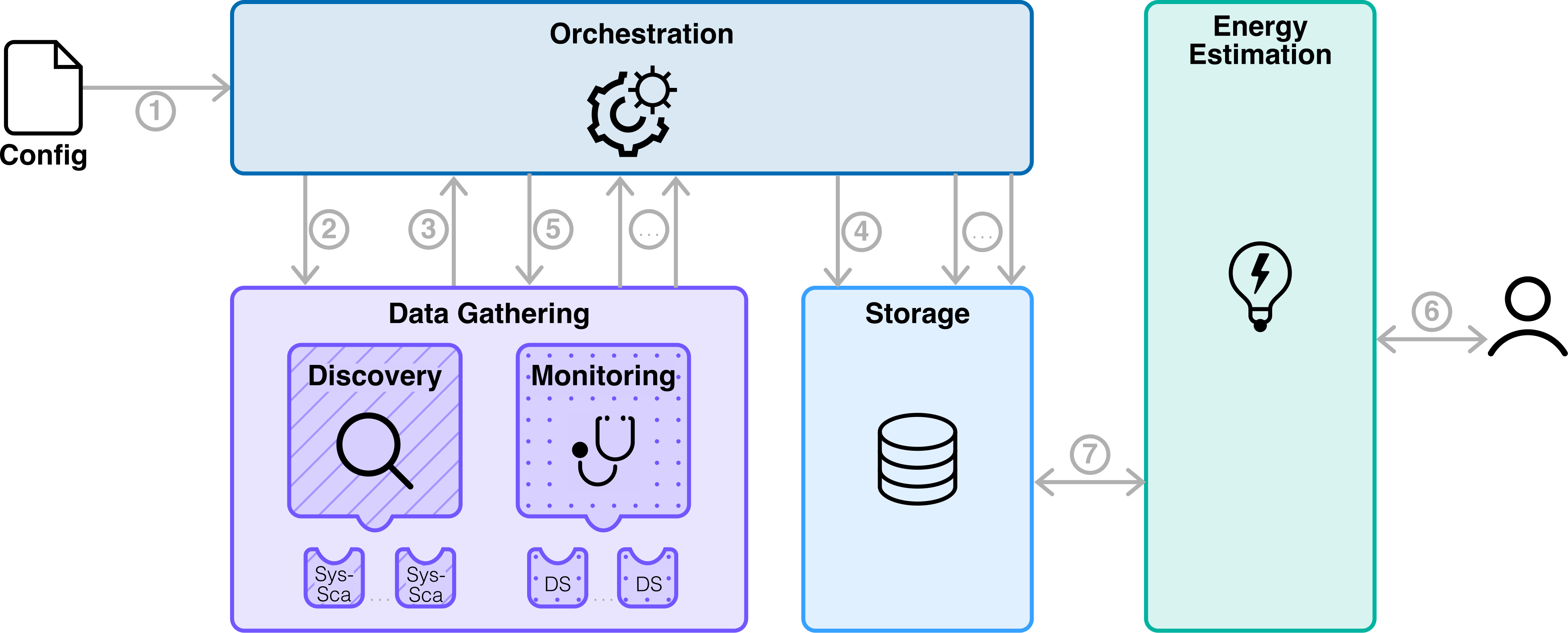}
\caption{Metrion Architecture and Dataflow}
\label{fig:Metrion}
\end{figure}

The METRION framework is divided into the data-gathering, storage, energy-attribution and orchestration components, connected through the PIDM, as shown in Figure \ref{fig:Metrion}.

The data gathering component is responsible for obtaining the necessary data points through its discovery and monitoring modules and transforming them into the platform-independent data representation required by the energy attribution model. The discovery module identifies and collects metadata about the physical components of the system and exposes these information as PhysicalEntity data objects. To deal with platform dependency, the discovery module offers the system scanning endpoint, through which platform-dependent discovery can be integrated. Hence, to support a new OS only a new dedicated system scanning functionality must be provided. The monitoring module obtains the measurements required by the energy attribution model through a set of data sources and provides them as Measurement data objects. To integrate heterogeneous data source backends for the monitoring module, a data source interface is provided. Each backend implementation specifies the physical resources and corresponding metrics it can collect. This design enables new data source backends to be added without modifying the core framework.

Next, the storage component stores the data, provided by the data gathering component, using a suitable storage backend. Since the data is consistently represented in this platform-independent format, the storage backend can be changed as needed without affecting the rest of the framework.

The energy attribution component forms the core of METRION. It implements the proposed energy attribution model to determine the energy consumption of the applications. To perform this attribution, the component retrieves the necessary measurement data from the storage component. Due to the platform-independent data model, the energy attribution process is rendered independent of platforms and hardware configurations. In addition, the component provides an endpoint that allows users to query the attributed energy consumption of a given application.

The orchestration component integrates the independent modules and coordinates the entire data and execution flow, as visualised in Figure \ref{fig:Metrion}. In the setup phase, the orchestrator first reads a configuration file that specifies parameters such as the monitoring time resolution and the selected storage backend \circnum{1}. It then calls the discovery module to collect metadata about the physical resources \circnum{2},\circnum{3} and stores this information using the storage component \circnum{4}. Based on the discovered hardware information, the orchestrator instructs the monitoring module to collect the measurements required by the energy assessment model \circnum{5}, which then uses the provided hardware information to select and configure the appropriate data sources. After the setup phase, the orchestrator regularly retrieves the collected metrics from the monitoring module \circnum{\tiny \dots} and forwards them to the storage component \circnum{\tiny \dots}. Finally, users can query the energy attribution component to obtain the attributed energy consumption values \circnum{6},\circnum{7}.

\subsection{Initial Implementation with Linux and Intel Extensions}
The METRION framework is primarily implemented in Python. The data-gathering and orchestration components are written entirely in Python, while the storage and energy attribution modules combine Python with SQL. The framework defines the system scanning and data source interfaces as abstract base classes to ensure consistent integration of new extensions.

Python is chosen for its portability across operating systems and its wide adoption, enabling easy extension of the framework with new system-scanning and data-source submodules. The storage component persists all measurements in an SQLite database, and the analysis module implements the EAM by combining Python logic with SQL queries. SQL is employed to leverage the database engine’s built-in optimization and aggregation capabilities. 

In order to support Linux-based systems with Intel CPUs featuring an SMT factor of two and two CPU packages, the METRION framework is extended with a system-scanning and two monitoring submodules.
The system-scanning component uses \linebreak /proc/cpuinfo and dmidecode to gather CPU and DRAM metadata. In addition, it determines the idle power of the CPU and DRAM by freezing all non-essential cgroups, after which the system remains idle for 60 seconds, and the average idle power of the CPU and DRAM is determined based on Intel RAPL readings.

To obtain the current energy consumption values of the CPU $E^{CPU}_{total}$ and DRAM $E^{DRAM}_{total}$ a data source for the Intel Powercap interface is implemented. In addition, an eBPF-based data source captures the required per-thread data points, i.e. $\Delta APERF$, $\Delta MPERF$, $\Delta UCC$, $\Delta Reads^{DRAM}_{local}$, and $Reads^{DRAM}_{remote}$, the core location, and the SMT factor. Therefore, the eBPF program attaches to the \linebreak sched\_switch tracepoint and then employs perf events to collect these metrics for each execution interval of a thread. Whether SMT is active during a thread’s execution interval is determined in post-processing. This approach is characterised by low-overhead, high-precision, and continuous sampling directly in the kernel and \linebreak thereby reducing the likelihood that the work of a thread is not captured. The implementation is publicly available on GitHub \footnote{https://github.com/ds-lab/Metrion}. 

\section{Evaluation}
The main goal of the proposed energy attribution model is to accurately capture the energy consumption of applications. To quantify its accuracy and assess its impact on system performance, the model is evaluated empirically across three dimensions.

The first dimension, attribution accuracy, measures how precisely the model attributes the active energy consumption of a workload on a component basis. Thereby, the accuracy is quantified with the Mean Absolute Percentage Error (MAPE) between the attributed value and a ground-truth reference. MAPE is chosen because its scale independence allows comparison between workloads and components with different energy ranges. The second dimension, stability, evaluates the consistency of the model’s attribution accuracy across three workloads, where the first workload primarily stresses the CPU, the second the memory, and the third both components. Thirdly, the dimension comparative accuracy compares the assessed accuracies to established tools. The following sections first describe the in-detail evaluation design, followed by the results and their discussion.
\begin{figure*}[ht!]
    \centering
    \begin{subfigure}[t]{0.32\textwidth}
        \centering
        \includegraphics[width=\linewidth]{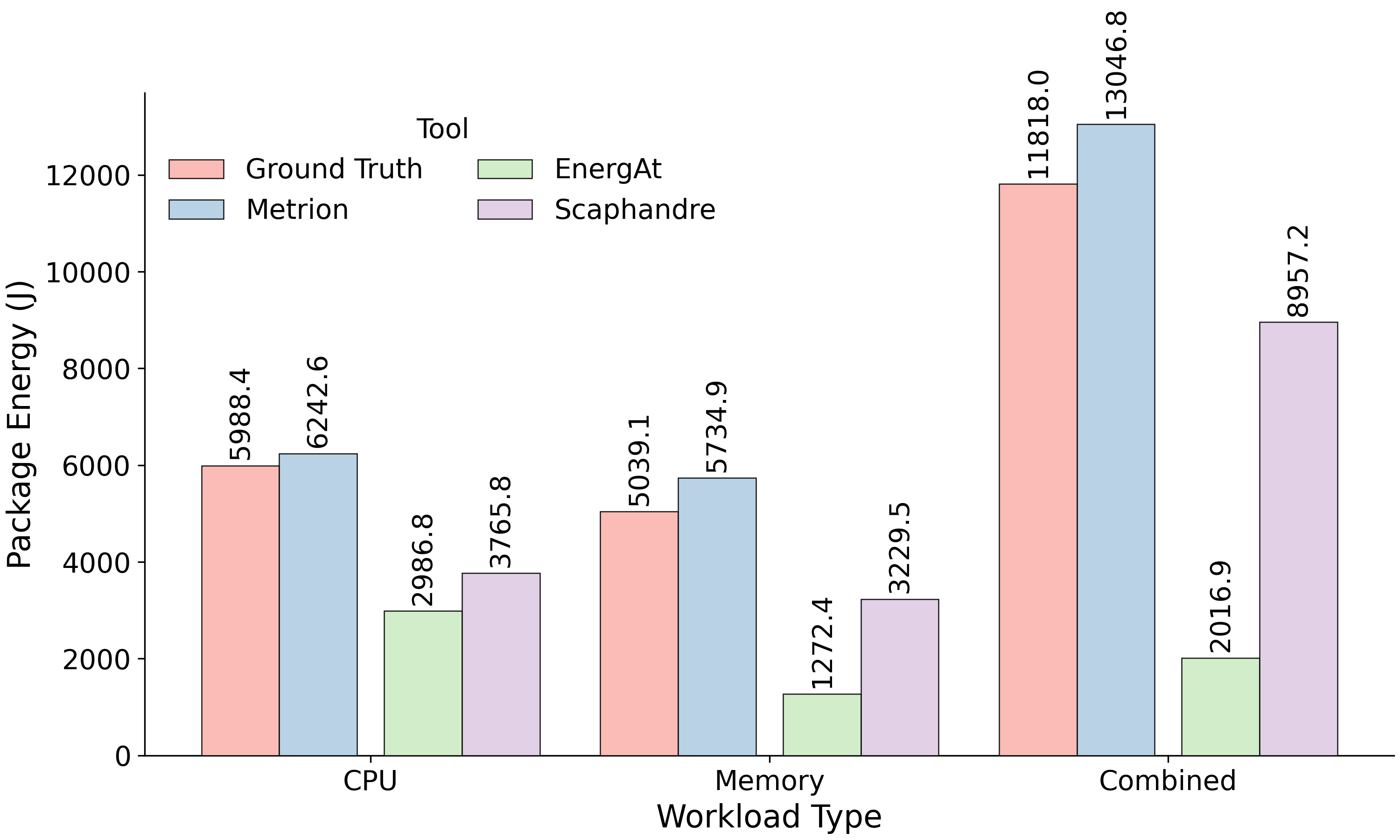}
        \caption{CPU Package Energy}
        \label{fig:setup_1_eval_pkg_energy}
    \end{subfigure}
    \hfill
    \begin{subfigure}[t]{0.32\textwidth}
        \centering
        \includegraphics[width=\linewidth]{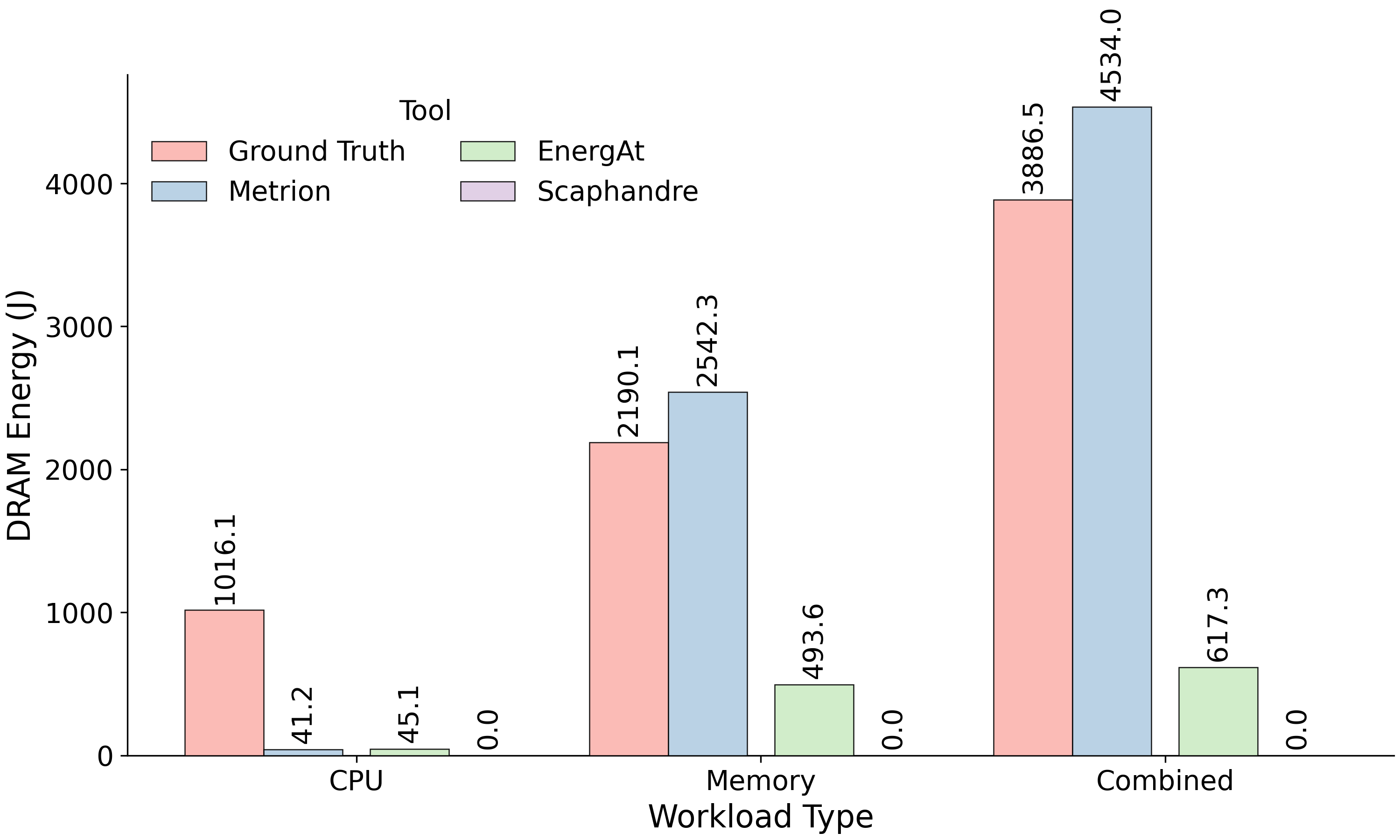}
        \caption{DRAM Energy}
        \label{fig:setup_1_eval_dram_energy}
    \end{subfigure}
    \begin{subfigure}[t]{0.32\textwidth}
        \centering
        \includegraphics[width=\linewidth]{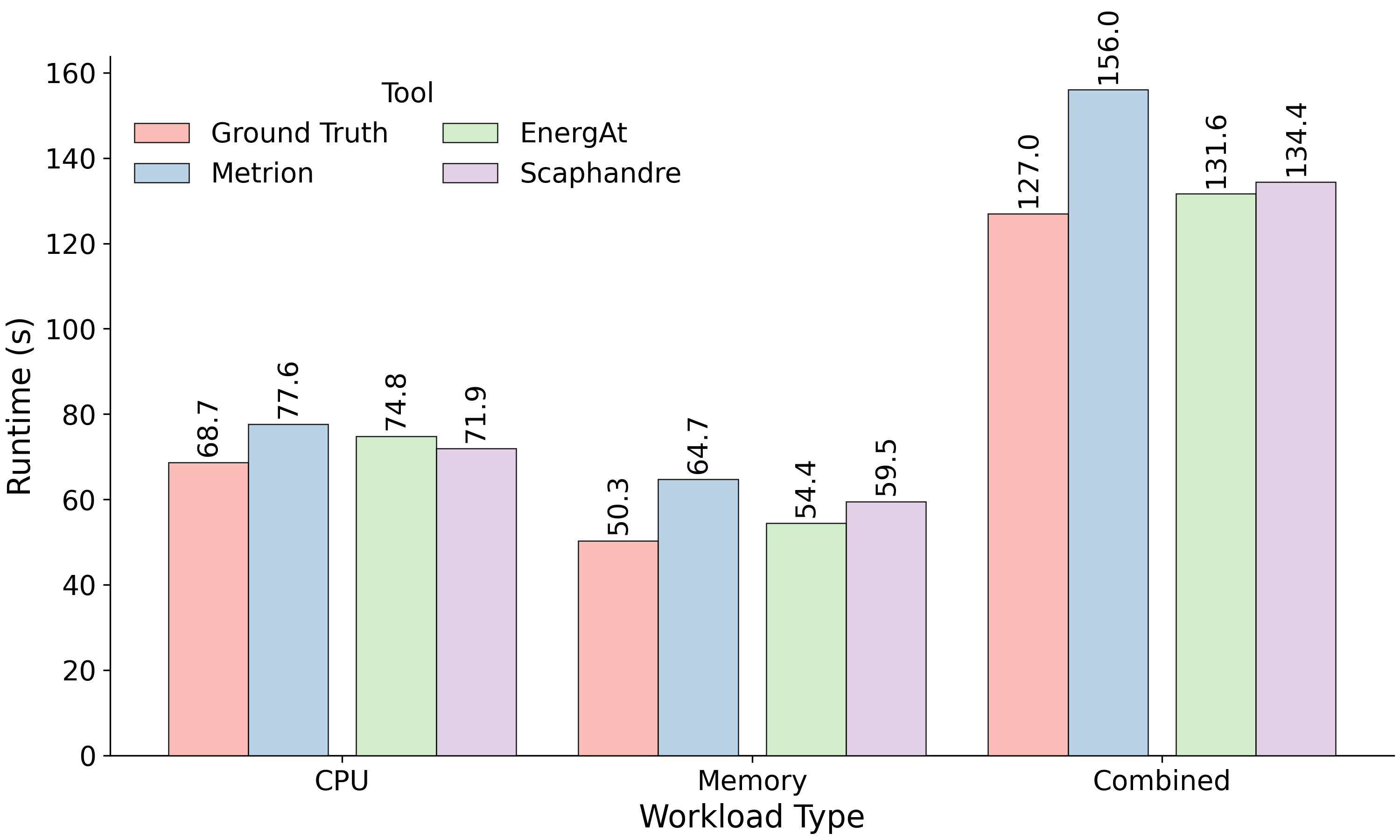}
        \caption{Runtime}
        \label{fig:setup_1_eval_runtime}
    \end{subfigure}
    
    \caption{CPU package energy, DRAM energy, and runtime results, each determined across three workloads: CPU-, Memory-, and Combined-workload.}
    \label{fig:eval_results}
\end{figure*}

\subsection{Evaluation Design}

To achieve reproducibility, the hardware environment, the experimental condition, and the ground-truth measurement procedure, as well as the selected workloads and energy measurement tools are described in detail. 

\subsubsection*{Hardware Setup:} All experiments are conducted three times on a system equipped with two Intel Xeon Silver 4110 CPUs, each with 8 physical cores and an SMT factor of two. The system is equipped with 384 GB of DDR4 ECC memory and runs Ubuntu 24.04.3 LTS with the Linux 6.8.0-85-generic kernel.

\subsubsection*{Experimental Condition:} To provide a controlled environment, all nonessential cgroups are frozen, and only the workload and a measurement tool are executed. Additionally, the CPU frequency governor is set to on-demand, and turbo mode is enabled to capture frequency-scaling effects.

\subsubsection*{Ground Truth Measurement:} The ground-truth reference energy \linebreak consumption is determined separately for each workload in two steps. In the first step, the idle power of each CPU and DRAM component is measured. Therefore, after all nonessential cgroups are frozen, the RAPL counters are read for each component, and the system is kept in this state for 60 seconds. Afterward, the RAPL counters are reread. The idle power consumption of each component is then determined by calculating the energy difference between two measured values and dividing it by the measurement interval. Following the idle power measurement, the active energy consumption for a given workload is determined in the second step. The nonessential cgroups remain frozen. The RAPL counters are read immediately before and after the workload execution, and the runtime of the workload is recorded. The difference between these readings gives the total energy consumption of the execution. To obtain the active energy, the idle energy is subtracted, which is calculated by multiplying the previously measured idle power by the workload’s runtime.

\subsubsection*{Workloads:} Three stressor from the stress-ng suite \cite{King.2017} are used to assess the stability of the energy attribution model. The \texttt{CPU workload} uses the cpu-method int32 stressor with $2,000,000$ operations to stress only the CPU, while the \texttt{Memory workload} uses the stream stressor with $200,000$ operations to specifically stress the memory. To stress both components simultaneously, the two stressors are combined to form the \texttt{Combined workload}.

\subsubsection*{Energy Measurement Tools:} For the third dimension, the comparative accuracy, the tools Scaphandre and EnergAt are selected, as these tools determine energy consumption at the application level. DEEP-Mon is not included in the comparison, as it is no longer actively maintained. Scaphandre reports the total CPU energy consumption, including the idle share, for each application, whereas EnergAt reports the active energy consumption of the CPU and DRAM components separately at the application level. The total EAM employed by Scaphandre is explicitly considered in the interpretation and discussion of the results.

\subsection{Results and Discussion}
The results are presented and discussed per physical component, and for each, the attribution accuracy, stability, and comparative accuracy are analyzed. Additionally, the overall system impact introduced by the measurement tools is examined.

\subsubsection{CPU Energy Attribution Accuracy} 
The proposed EAM determines CPU energy consumption for workloads with relatively high accuracy, with an average MAPE of 9.5\% across all three workloads. For comparison, EnergAt and Scaphandre achieve average MAPEs of 69.3\% and 32.4\%, respectively. The model performs best for the \texttt{CPU workload}, with an MAPE of 4.2\%, and worst for the \texttt{Memory workload}, with an MAPE of 13.8\%, as shown in Figure \ref{fig:setup_1_eval_pkg_energy}. The low error for the \texttt{CPU workload} and the higher error for the \texttt{Memory workload} may be due to the design of the RAPL package's power domain and the model's focus on core components. The package power domain includes the core components and uncore components, such as the last-level cache. In the case of the \texttt{CPU workload}, mainly the core components are stressed, whereas for the \texttt{Memory workload}, both the core and uncore components are utilized. As the proposed model does not explicitly model the uncore activity by a workload, a higher error is introduced. In contrast, both other tools determine the CPU energy consumption considerably lower across all three workloads, including Scaphandre, even though it reports total energy consumption. Nevertheless, Scaphandre captures the overall trend of the ground-truth energy consumption and shows that when the ground-truth increases, as in the case of the \texttt{Combined workload}, the reported value also increases, whereas for the \texttt{Memory workload}, both values decrease. EnergAt, however, does not capture this trend. Overall, the proposed model accurately captures CPU energy consumption, and its standard deviation of 3.95\% and coefficient of variation of 4.17\% indicate stable performance across different workload types.

\subsubsection{DRAM Energy Attribution Accuracy} 
The DRAM energy consumption of a workload is partially accurately determined by the proposed EAM across different workloads. For the \texttt{CPU workload}, the model underestimates DRAM energy consumption, with an MAPE of 95.6\%. This may result from CPU operations targeting DRAM that do not correlate with read operations as modeled by the EAM. In contrast, for the \texttt{Memory workload} and \texttt{Combined workload}, the MAPE decreases notably to around 16\%, as shown in Figure \ref{fig:setup_1_eval_dram_energy}. This improvement may occur because these workloads stress the memory through read operations captured by the model, while the remaining error likely arises from unmodeled write operations. Only EnergAt reports per-application DRAM energy consumption and captures the overall trend across workloads, but still determines absolute values considerably lower, with an average MAPE of ~85.7\%. Overall, the EAM requires refinement, as it captures DRAM energy consumption with mixed accuracy.
 
\subsubsection{System Impact Analysis}
All three tools affect system performance, as their monitoring increases the total runtime of each workload. However, the impact varies with the workload type, with the \texttt{Memory workload} showing the most significant impact and the CPU workload the least, as shown in Figure \ref{fig:setup_1_eval_runtime}. Metrion causes the highest average increase in runtime across all three workloads at 21.1\%, followed by Scaphandre at 9.2\% and EnergAt at 6.5\%. The different effects on system performance may be caused by the different implementation languages used. Metrion is mainly implemented in Python, Scaphandre in Rust, and EnergAt uses C/C++ for performance-critical components. To reduce Metrion's higher overhead, performance-critical parts could be implemented in Rust.

\section{Conclusion}
Accurately determining the energy consumption of applications is required to identify optimization potential at both the macro- and micro-architectural levels and to quantify the improvements achieved by optimization. To achieve this, we introduced an EAM that accounts for the influence of SMT, frequency scaling, multi-socket architectures, and NUMA on the energy consumption of an application at the thread level. In order to use the EAM across different platforms and at both the micro- and macro-architectural levels, we propose the extensible framework METRION, along with a platform-independent data model. Additionally, we provided an initial implementation of METRION for Linux-based systems using Intel CPUs. The evaluation demonstrates that the EAM can accurately capture the CPU energy consumption of applications targeting solely the CPU with a MAPE of 4.2\%, and applications that target the DRAM explicitly with a MAPE of 16.1\%. However, the framework increases the runtime of workloads, introducing a system overhead.
In further work, we aim to improve the EAM and the METRION framework. Firstly, we plan to refine the CPU model by accounting for uncore activity, and the DRAM model by capturing DRAM write activity. Additionally, we intend to extend the overall model to capture energy consumption resulting from network activity and GPU usage. Secondly, we plan to reduce the system impact of METRION by rewriting performance-critical components in Rust. Thirdly, we plan to evaluate METRION on additional hardware platforms, such as AMD-based systems, and to systematically study how varying the number of executed workload operations influences attribution accuracy. Overall, the proposed EAM and framework lay the foundation to identify and quantify optimisations across micro- and macro-architectural levels, such as within the edge–fog-cloud continuum.

\begin{acks}
\textbf{Funding.} 
This research was supported by the Federal Ministry of Research, Technology and Space under grant number 16IS24070G.

\textbf{Transparency note.} 
The readability of this text has been improved with the support of AI technologies.
\end{acks}
\bibliography{ref}
\end{document}